\documentclass[aps,twocolumn,pra,letterpaper,superscriptaddress,nofootinbib,longbibliography]{revtex4-1}
\usepackage{amssymb}
\usepackage{physics}
\usepackage{mathtools}
\usepackage{upgreek}
\usepackage{booktabs}
\usepackage{txfonts}
\usepackage[pdftex]{graphicx}
\usepackage[usenames, dvipsnames, svgnames, table]{xcolor}
\usepackage[colorlinks=true, citecolor=Blue,linkcolor=BrickRed, urlcolor=magenta]{hyperref}
\usepackage{bm}
\usepackage{placeins} 
\usepackage{setspace}
\usepackage[final]{changes}
\definechangesauthor[color=purple, name={Miguel Dovale}]{MD}

\begin{document}


\newcommand{\R}[1]{\textcolor{red}{#1}}
\newcommand{\B}[1]{\textcolor{blue}{#1}}
\newcommand{\finesse}{\mathcal{F}}
\def\*#1{\bm{#1}}
\def\fw#1{\texttt{#1}}
\newcommand{\RXG}{MBF} 
\newcommand{\MBF}{MBF} 
\newcommand{\GMF}{GMF} 
\newcommand{\Rx}{Rx} 
\newcommand{\Tx}{Tx} 
\newcommand{\LO}{LO} 

\newcommand{\SHS}{SHS}  
\newcommand{\refSEPD}{refSEPD}
\newcommand{\refQPD}{refQPD}
\newcommand{\tempSEPD}{tempSEPD}
\newcommand{\SciQPD}{SciQPD}


\title{Optical suppression of tilt-to-length coupling in the LISA long-arm interferometer}

\author{M~Chwalla}
\affiliation{Airbus DS GmbH, Claude-Dornier-Stra{\ss}e, 88090 Immenstaad, Germany}

\author{K~Danzmann}
\affiliation{Max Planck Institute for Gravitational Physics (Albert Einstein Institute) and Institute for Gravitational Physics of the Leibniz Universit\"at Hannover, Callinstra{\ss}e~38, 30167 Hannover, Germany}

\author{M~{Dovale~\'Alvarez}}
\email{miguel.dovale@aei.mpg.de}
\affiliation{Max Planck Institute for Gravitational Physics (Albert Einstein Institute) and Institute for Gravitational Physics of the Leibniz Universit\"at Hannover, Callinstra{\ss}e~38, 30167 Hannover, Germany}

\author{J~J~{Esteban~Delgado}}
\affiliation{Max Planck Institute for Gravitational Physics (Albert Einstein Institute) and Institute for Gravitational Physics of the Leibniz Universit\"at Hannover, Callinstra{\ss}e~38, 30167 Hannover, Germany}

\author{G~{Fern\'andez~Barranco}}
\affiliation{Max Planck Institute for Gravitational Physics (Albert Einstein Institute) and Institute for Gravitational Physics of the Leibniz Universit\"at Hannover, Callinstra{\ss}e~38, 30167 Hannover, Germany}

\author{E~Fitzsimons}
\affiliation{Airbus DS GmbH, Claude-Dornier-Stra{\ss}e, 88090 Immenstaad, Germany}
\affiliation{UK Astronomy Technology Centre, Royal Observatory Edinburgh, Blackford Hill, Edinburgh EH9 3HJ, UK}

\author{O~Gerberding}
\affiliation{Max Planck Institute for Gravitational Physics (Albert Einstein Institute) and Institute for Gravitational Physics of the Leibniz Universit\"at Hannover, Callinstra{\ss}e~38, 30167 Hannover, Germany}
\affiliation{Institute for Experimental Physics, University of Hamburg, Luruper Chaussee 149, 22761, Hamburg, Germany}

\author{G~Heinzel}
\email{gerhard.heinzel@aei.mpg.de}
\affiliation{Max Planck Institute for Gravitational Physics (Albert Einstein Institute) and Institute for Gravitational Physics of the Leibniz Universit\"at Hannover, Callinstra{\ss}e~38, 30167 Hannover, Germany}

\author{C~J~Killow}
\affiliation{SUPA, Institute for Gravitational Research, University of Glasgow, Glasgow G12 8QQ, Scotland, UK}

\author{M~Lieser}
\affiliation{Max Planck Institute for Gravitational Physics (Albert Einstein Institute) and Institute for Gravitational Physics of the Leibniz Universit\"at Hannover, Callinstra{\ss}e~38, 30167 Hannover, Germany}

\author{M~Perreur-Lloyd}
\affiliation{SUPA, Institute for Gravitational Research, University of Glasgow, Glasgow G12 8QQ, Scotland, UK}

\author{D~I~Robertson}
\affiliation{SUPA, Institute for Gravitational Research, University of Glasgow, Glasgow G12 8QQ, Scotland, UK}

\author{J~M~Rohr}
\affiliation{Max Planck Institute for Gravitational Physics (Albert Einstein Institute) and Institute for Gravitational Physics of the Leibniz Universit\"at Hannover, Callinstra{\ss}e~38, 30167 Hannover, Germany}

\author{S~Schuster}
\affiliation{Max Planck Institute for Gravitational Physics (Albert Einstein Institute) and Institute for Gravitational Physics of the Leibniz Universit\"at Hannover, Callinstra{\ss}e~38, 30167 Hannover, Germany}

\author{T~S~Schwarze}
\affiliation{Max Planck Institute for Gravitational Physics (Albert Einstein Institute) and Institute for Gravitational Physics of the Leibniz Universit\"at Hannover, Callinstra{\ss}e~38, 30167 Hannover, Germany}

\author{M~Tr\"obs}
\affiliation{Max Planck Institute for Gravitational Physics (Albert Einstein Institute) and Institute for Gravitational Physics of the Leibniz Universit\"at Hannover, Callinstra{\ss}e~38, 30167 Hannover, Germany}

\author{G~Wanner}
\email{gudrun.wanner@aei.mpg.de}
\affiliation{Max Planck Institute for Gravitational Physics (Albert Einstein Institute) and Institute for Gravitational Physics of the Leibniz Universit\"at Hannover, Callinstra{\ss}e~38, 30167 Hannover, Germany}

\author{H~Ward}
\affiliation{SUPA, Institute for Gravitational Research, University of Glasgow, Glasgow G12 8QQ, Scotland, UK}

\begin{abstract}

The arm length and the isolation in space enable LISA to probe for signals unattainable on ground, opening a window to the sub-Hz gravitational-wave universe. The coupling of unavoidable angular spacecraft jitter into the longitudinal displacement measurement, an effect known as tilt-to-length (TTL) coupling, is critical for realizing the required sensitivity of picometer$/\sqrt{\rm{Hz}}$. An ultra-stable interferometer testbed has been developed in order to investigate this issue and validate mitigation strategies in a setup representative of LISA, \added{and in this paper it is operated in the long-arm interferometer configuration}. \added{The testbed is fitted with a flat-top beam generator to simulate the beam received by a LISA spacecraft. We demonstrate a reduction of TTL coupling between this flat-top beam and a Gaussian reference beam via introducing two- and four-lens imaging systems.} TTL coupling factors below $\pm 25\,\mu$m/rad for beam tilts within $\pm 300\,\mu$rad are obtained by careful optimization of the system. Moreover we show that the additional TTL coupling due to lateral alignment errors of elements of the imaging system can be compensated by introducing lateral shifts of the detector, and vice versa. These findings help validate the suitability of this noise-reduction technique for the LISA long-arm interferometer.

\end{abstract}

\maketitle

\section{Introduction}
	
\begin{figure*}
\centering
\includegraphics[scale=0.85]{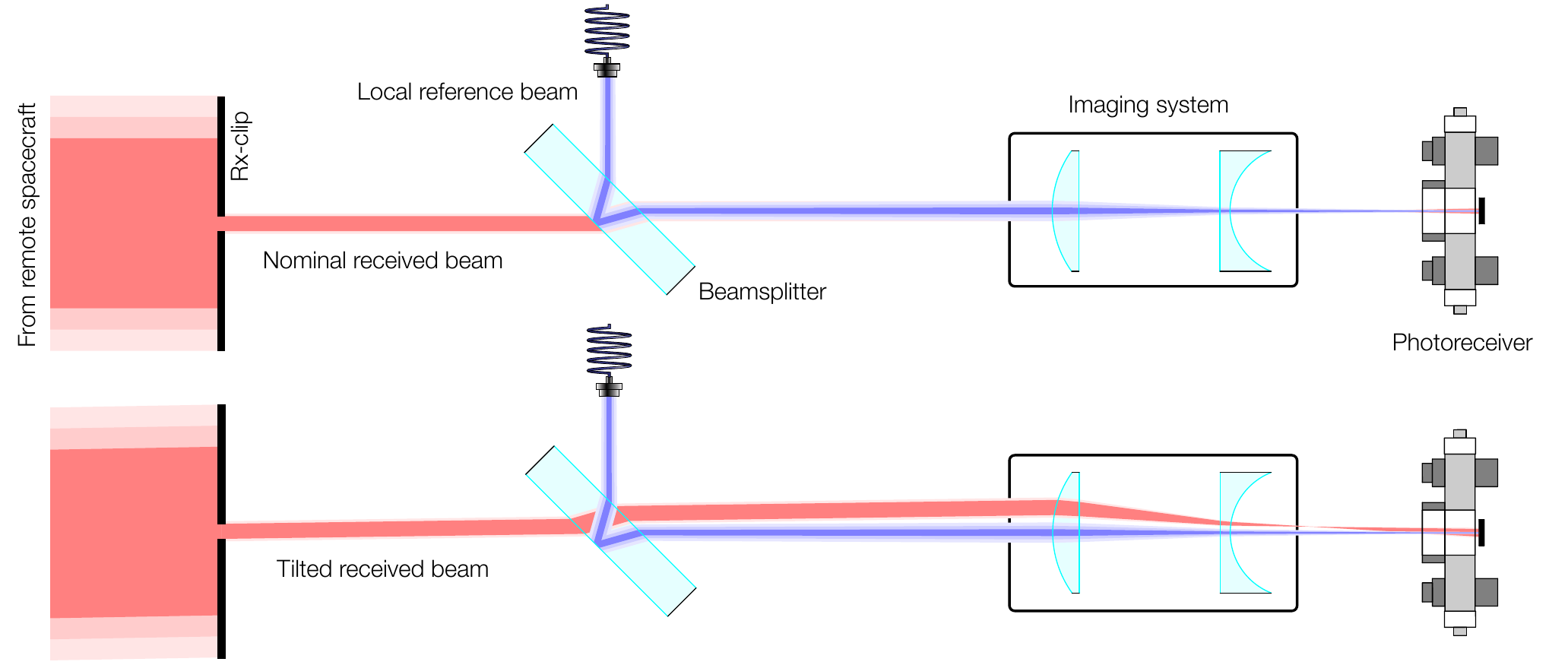}
\caption{Tilt-to-length coupling in the LISA long-arm interferometer. \emph{Top:} The received beam, that propagated through the $\sim 2.5\cdot 10^{9}$~m-long arm, is captured by a telescope and imaged onto the \Rx{}-clip, a small aperture located in the optical bench. The received light, which has a flat-top profile as a result of clipping the large beam at a small circular aperture, interferes with a locally generated Gaussian beam, and their heterodyne beat note --- encoding the gravitational-wave signal --- is registered by a quadrant photodiode. \emph{Bottom:} Angular jitter of the spacecraft causes a tilt of the \Rx{} beam with respect to the \Rx{}-clip, and a consequent error in the longitudinal pathlength signal measurement, a source of noise known as tilt-to-length coupling (the tilt has been exaggerated in the figure for illustration). Imaging systems can be used to image the \Rx{}-clip onto the detector plane, hence reducing TTL coupling in the interferometer and increasing LISA's robustness to alignment noise.}
\label{figure:concept}
\end{figure*}

Directly sensing gravitational effects by tracking the motion between freely moving masses is at the center of experimental gravitational physics, with various exciting results achieved in recent years. On 14 September 2015 the Laser Interferometer Gravitational-Wave Observatory (LIGO)~\cite{Collaboration2015}, consisting of two second generation detectors listening in the Hz to kHz band, made the first direct observation of gravitational waves from a binary black hole merger (GW150914)~\cite{GW150914}. The discovery consolidated laser interferometry as a suitable technology for gravitational-wave detection, and strengthened the revolutionary scientific value and discovery potential of a deep-space gravitational-wave observatory capable of listening to sub-Hz gravitational-wave signals, such as the Laser Interferometer Space Antenna (LISA)~\cite{elisa13ARXIV, LISAMissionProposal17ARXIV}.

On 3 December 2015 the European Space Agency (ESA) launched LISA Pathfinder (LPF), a single satellite technology demonstrator for LISA, providing successful flight demonstration of critical LISA-like instruments~\cite{Armano2016, armano2018beyond, Wanner2019}. On 22 May 2018, the National Aeronautics and Space Administration (NASA) and the German Research Centre for Geosciences (GFZ) launched GRACE Follow-On (GRACE-FO), a twin satellite gravity exploration mission carrying a laser ranging instrument, successfully proving LISA-like technologies for inter-spacecraft optical links, albeit on a much smaller baseline of only 200\,km~\cite{GFO2019}.

LISA has been approved by ESA as the ESA-L3 Gravitational Wave Mission~\cite{LISA2017}, and while it will feature flight-proven instruments and interferometry techniques, it also presents many new technical challenges. The LISA long-arm interferometer features the longest baseline of any human-made gravity and spacetime experiment in history with approximately $2.5$ million kilometers. LISA will be able to probe a large and as-yet unexplored part of the \emph{gravitational universe}, measuring gravitational waves in a frequency band (0.1\,mHz to 1\,Hz) where current ground-based laser interferometers, such as LIGO or Virgo, have poor sensitivity~\cite{Martynov2016}. The unprecedented laser interferometric baseline in combination with the desired precision, however, presents unique challenges that can be considered fundamental for such measurement schemes.

The LISA satellites exchange laser signals that probe the displacement between the satellites' free-floating test masses in order to detect spacetime strain between the spacecraft. A locally generated Gaussian beam is emitted via a telescope, propagates through the $\sim 2.5\cdot 10^{9}$~m-long interferometer arm, and is captured by a similar telescope in the remote spacecraft. The received light, which is the result of clipping the center region of a kilometer-scale light beam with a small aperture, is interfered with the local beam, and their heterodyne beat is used to obtain a measurement of the length fluctuations of the arms with picometer$/\sqrt{\text{Hz}}$ precision, encoding the gravitational-wave signal.

This measurement is challenged by many noise sources at the local instrument level, at the optical link level, and of residual origin. Tilt-to-length coupling is one of the most significant contributions to the total noise budget in LISA (and also in LPF~\cite{Wanner17JPCS} and GRACE-FO~\cite{Sheard12JGeod, Wegener2020}). Relative angular motion between the test mass and spacecraft, and between remote spacecraft, results in angular jitter of the interfering beams and a consequent error in the longitudinal displacement measurement. 


In LISA, TTL coupling noise is expected to be very significant, with a large contribution stemming from the long arm interferometer, partly due to the magnification provided by the telescope, which increases the tilt sensed by the photoreceivers. Therefore, a system for suppressing this cross-talk is essential. One such approach is to use imaging systems to image the tilting beams onto the photoreceivers. In the long-arm interferometer, angular motion of the spacecraft translates into tilt of the measurement beam with respect to a fixed  aperture on the optical bench, the \Rx{}-clip. Thus, an optical system configured to image the \Rx{}-clip onto the detectors can significantly lower the impact of this noise source in the interferometer (Figure~\ref{figure:concept}).

In this paper we present an investigation of TTL coupling noise reduction in the LISA long-arm interferometer via two- and four-lens imaging systems. A testbed has been developed which incorporates a pair of ultra-stable Zerodur interferometer platforms. The testbed has been used previously \added{in another configuration} to demonstrate TTL coupling noise reduction in the test mass interferometer~\cite{Trobs2018}. The long-arm interferometer, however, presents a unique challenge due to the nature of the received light, which has a flat amplitude and phase profile~\cite{Waluschka1999, Papalexandris2003, Sasso2018, Sasso2018b}. 

\added{In order to simulate the received beam by a LISA spacecraft, the testbed is fitted with a flat-top beam generator, which produces a beam with flat intensity and phase profiles. This allows us to carry out an interferometric measurement using a tilting flat-top beam, which undergoes diffraction at the \Rx{}-clip and is imaged onto the detector, and a reference Gaussian beam, a situation representative of the LISA long-arm interferometer. We demonstrate that imaging systems meet the required performance of TTL coupling factors below $\pm 25$\,$\mu$m/rad (i.e., $\pm 1\,$pm/40\,nrad) for beam tilt angles within $\pm 300\,\mu$rad, a performance requirement derived from a top-level breakdown carried out in a previous mission study~\cite{Weise10TN6}. Furthermore, we demonstrate that the residual TTL coupling in the interferometer with imaging systems can be counteracted by intentionally introducing lateral offsets of the components of the system.}

Section~\ref{section:testbed} presents the layout of the testbed and relevant subsystems, \added{and describes the performance of the flat-top beam generator}. An introduction to tilt-to-length coupling noise, and to the suppression mechanism investigated in this paper, is given in Section~\ref{section:imaging-systems}. The experimental results are provided in Section~\ref{section:results}. \added{We show the achieved performance of the imaging systems for TTL coupling suppression, and detail the additional experimental effort that was necessary in order to obtain these results}. We investigate the robustness of the setup against alignment errors of the system, and present a tolerance analysis. \added{Finally, we show how the residual TTL coupling of the system can be compensated by component alignment.}

\begin{figure*}
\centering
\includegraphics[scale=0.85]{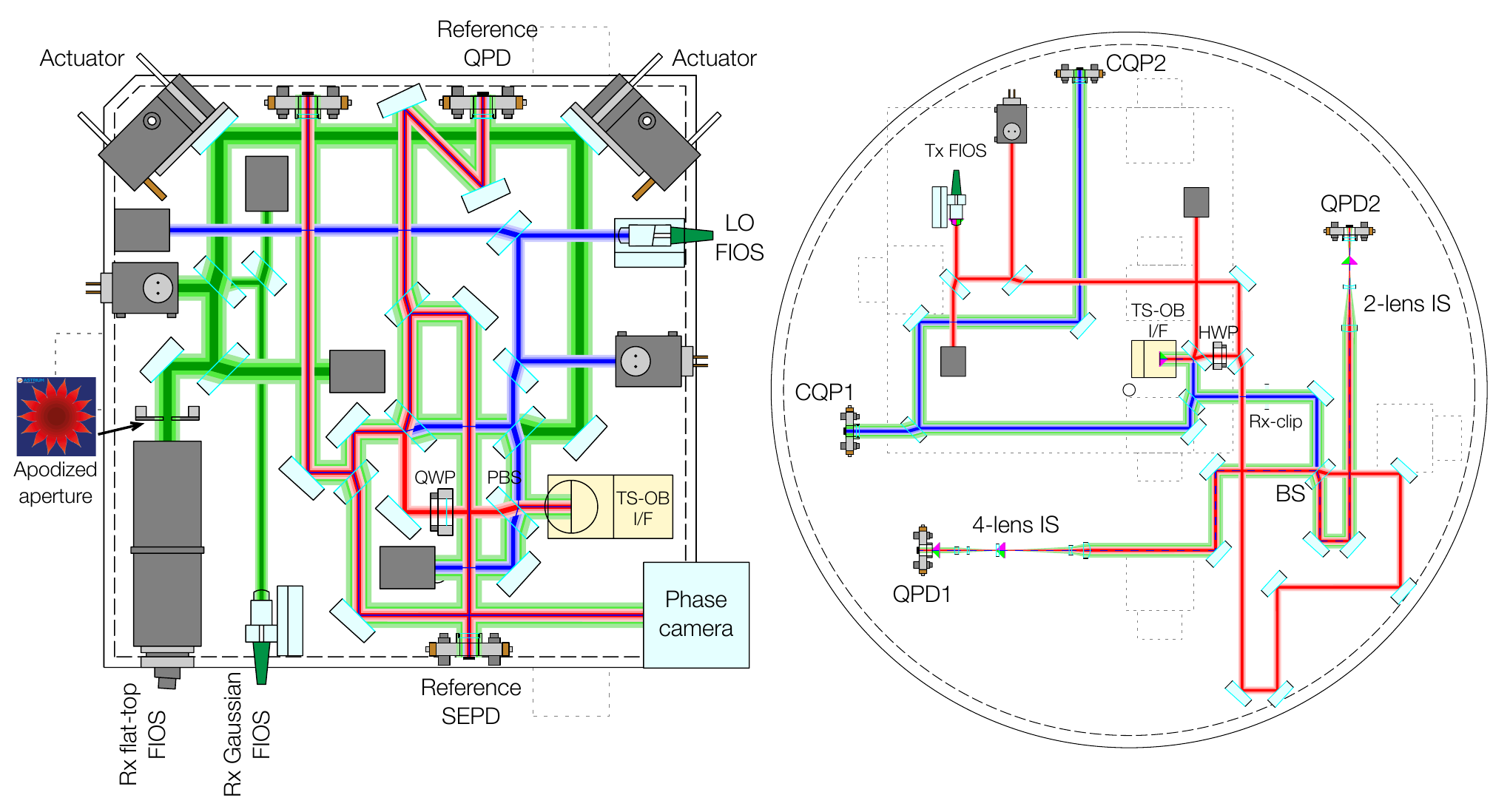}
\caption{\added{Layout of the testbed's telescope simulator (left) and optical bench (right). All beams are prepared from a single laser source by splitting the beam and passing through three acousto-optic modulators (not shown) before being injected into the testbed through the fiber injector optical subassemblies (FIOS) depicted here. \emph{Left:} The telescope simulator (TS) provides the optical bench (OB) with laser beams to simulate the LISA long-arm (LA) and test mass (TM) interferometers. To simulate the LA interferometer, the \Rx{} flat-top beam is generated using a commercial fiber collimator and a custom-made apodization aperture (shown on the lower left corner). The TS seats on the surface of the OB using a three-point support system with Zerodur feet, and an optical link is established via a vertical interface (TS-OB I/F). Actuated mirrors on the TS tilt the \Rx{} beam with respect to the \Rx{}-clip to simulate angular jitter of the spacecraft. \emph{Right:} In the OB, the \Rx{} beam from the TS interferes with the locally generated \Tx{} beam, and the resulting heterodyne beat notes are read out using a balanced detection scheme with quadrant photodiodes (QPD1 and QPD2) at the output ports of the recombination beamsplitter (BS). Tilts of the \Rx{} beam translate to apparent longitudinal motion of the spacecraft, an effect known as tilt-to-length coupling. This source of noise is greatly reduced by placing specially designed imaging systems (IS) before the measurement QPD's. An additional local oscillator beam (\LO{}) is also generated in the TS to aid the TS-OB alignment and calibration. A series of auxiliary interferometers are required for calibration of the experiment: a calibrated quadrant photodiode pair (CQP1 and CQP2) is used for alignment of the TS relative to the OB; the reference single element photodiode (SEPD) in the TS, positioned as an optical copy of the \Rx{}-clip, is used to offset-phase-lock the \Rx{} and \Tx{} beams to the \LO{} beam, ensuring that TTL coupling originating in the TS is rejected; finally, the calibrated differential wavefront sensing (DWS) signals from the reference QPD in the TS are used to obtain the tilt angle of the \Rx{} beam.}}
\label{figure:setup}
\end{figure*}

\section{Experimental Methods}
\label{section:testbed}

\subsection{LISA Optical Bench Testbed}

The LISA Optical Bench (LOB)~\cite{dArcio2017}, of which there are two per spacecraft, will likely consist on an ultra low expansion glass baseplate to which other optical elements are bonded via hydroxide-catalysis bonding~\cite{HC_Bonding2014}. The LOB hosts critical components of the different interferometers that make up the LISA detector. The long-arm (LA) interferometer tracks the relative length fluctuations between satellites by measuring the relative phase between the long-traveled received beam (\Rx{}) and a locally generated beam (\Tx{}), as well as the relative alignment via the differential wavefront sensing (DWS) signals~\cite{Morrison1994, Hechenblaikner10}. The test mass (TM) interferometer measures the longitudinal displacement between the test mass and the LOB using the \Tx{} beam and another locally generated beam (\LO{}), as well as the tilt of the test mass via DWS. Finally, the LOB is fitted with an additional interferometer providing the phase reference between the two LOBs on one spacecraft. The measurement of the relative displacement between two distant test masses, which carries the gravitational-wave signal, is obtained by combining three contributions: the length fluctuations between the distant LOBs, and the length fluctuations between each LOB and their respective test mass.

Any effect causing alignment errors between the beams in the LA or TM interferometers (e.g., angular jitter of the spacecraft or test masses) will decrease the sensitivity in the gravitational-wave signal channel via cross-coupling of beam tilt to apparent longitudinal motion. For example, in the LA interferometer the wavefront captured by the telescope is nearly flat, but it tilts with respect to the entrance pupil of the telescope due to spacecraft angular motion. The telescope compresses the beam, imaging it onto an aperture called the \Rx{}-clip placed on the optical bench. Therefore, angular jitter of the spacecraft manifests as tilting of the \Rx{} beam with respect to the \Rx{}-clip. A testbed for simulating these effects in the LOB has been developed at the AEI~\cite{Chwalla16CQG, Trobs2018}, consisting of a cylindrical Zerodur optical bench (OB) and a rectangular Zerodur telescope simulator (TS), as shown in Figure~\ref{figure:setup}.

The OB \added{(Figure~\ref{figure:setup} right)}, a 55\,kg 580\,mm-diameter 80\,mm-thick Zerodur cylinder, is a simplified version of the LOB, containing only the relevant components for our investigation. The main measurement interferometer in the OB consists on a single recombination beamsplitter \added{(labeled BS in Figure~\ref{figure:setup} right)} with a balanced detection scheme, using a pair of quadrant photodiodes (QPD) for reading out the heterodyne beat notes of either the LA or TM interferometers, depending on the input from the TS \added{(i.e., the \Rx{} flat-top or \Rx{} Gaussian beams, respectively)}. This interferometer has a nominal optical pathlength mismatch of zero between the arms in order to minimize coupling of laser frequency noise into displacement noise. An auxiliary interferometer using a pair of calibrated quadrant photodiode pairs \added{(CQP1 and CQP2)} is also fitted for aiding alignment and calibration of the TS.

The TS provides the OB with input beams that simulate the TM and the LA interferometers, and induces angular deviations of the beams (\Rx{}) via actuation of a pair of motorized steering mirrors. These mirrors are mounted on a custom-designed mount that suppresses any thermally induced displacements of the steering mirrors. The TS assembly, consisting of a $280\,$mm$\,\times\,280\,$mm rectangular Zerodur block mounted on a three-point support structure with Zerodur feet, rests on the surface of the OB, and an optical link is established between the two subsystems via a vertical interface. The TS also supplies the \LO{} beam, which does not interact with the actuated mirrors and is used for aligning the TS to the OB.

The design of the TS (Figure~\ref{figure:setup} left) guarantees that, when it is correctly linked to the OB, any actuation of the steering mirrors causes a tilt of the \Rx{} beam around the \Rx{}-clip in the OB (Figure~\ref{figure:setup} right). An optical phase-locked loop (OPLL) between the \LO{} and \Rx{} beams, and another between the \LO{} and \Tx{} beams, \added{using the phase signal of the reference single-element photodiode (SEPD) on the TS,} ensures that the relative phase between the \Rx{} and \Tx{} beams at the \Rx{}-clip is zero. \added{To this end the TS is fitted with a reference interferometer (Reference SEPD) measuring the phase of all three beat notes in a position that is an optical copy of the \Rx{}-clip} with respect to the \Rx{} beam (i.e., the \Rx{} beam experiences the same phase change and motion, and has the same geometry, at both locations). This is essential for ensuring that the TTL coupling measurement is free from any contributions from the TS, and only the TTL coupling that is intrinsic to the optical bench is observed.

\begin{figure}
\centering
\includegraphics[scale=0.85]{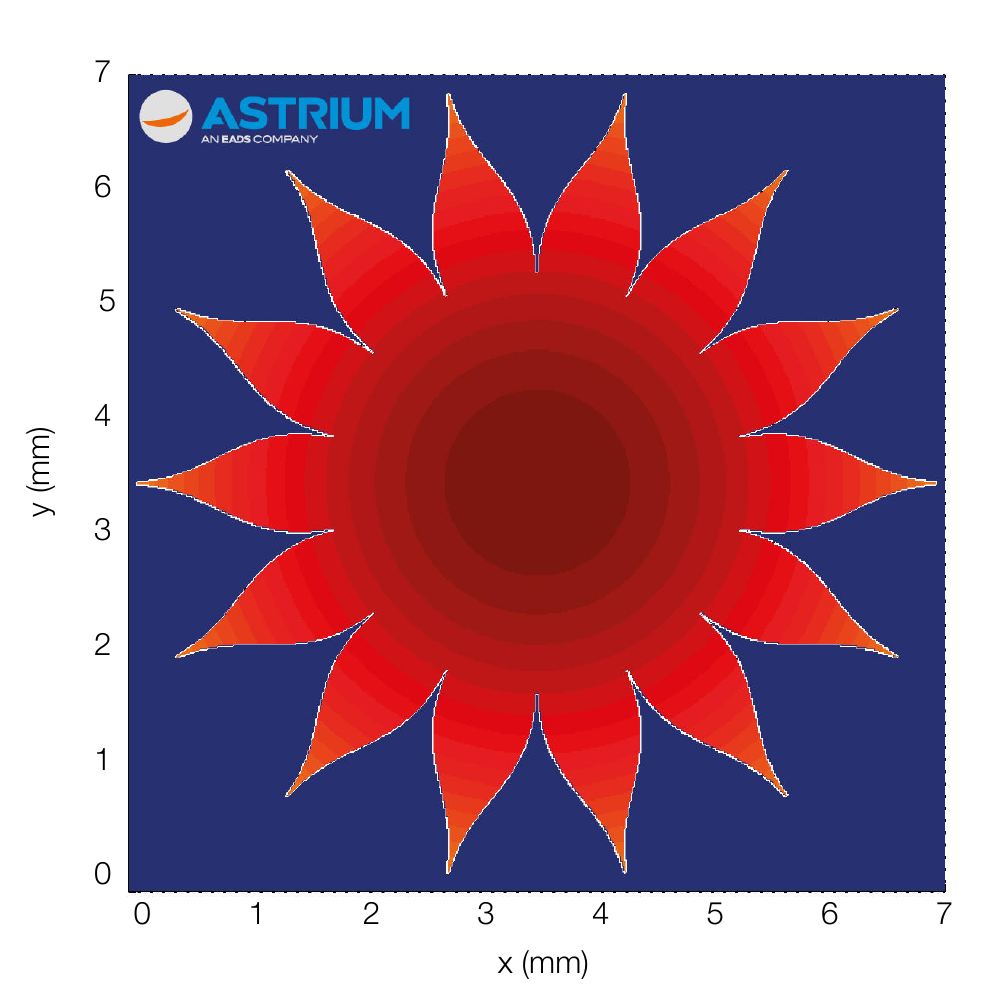}
\caption{\added{Apodization aperture for flat-top beam generation. To generate a beam representative of the received beam in the LISA long-arm interferometer, a collimated 9\,mm Gaussian beam, clipped to a diameter of 24\,mm, is diffracted through an apodization aperture of the shape shown. The ``starshade'' shape, which yields an offset hyper-Gaussian transmission function, is obtained via a Monte Carlo simulation optimizing the flatness of the diffracted beam pattern at the position of the \Rx{}-clip. The result is a flat-top beam with $\lambda$/100 PV nominal phase flatness.}}
\label{figure:APO-aperture}
\end{figure}


\subsection{Flat-top beam generator}

\added{To simulate the LA interferometer, the TS produces a flat-top beam representative of the far-field beam received from the remote spacecraft. For this purpose the TS is fitted with a commercially available fiber collimator by Sch\"after \& Kirchhoff (60FC-T-4-M100S-37) that provides a 9\,mm-radius Gaussian beam clipped to a diameter of 24\,mm. This beam then diffracts at a custom-designed apodization aperture mounted at the immediate output of the collimator, which eliminates diffraction rings in the far-field that are due to the clipping at the collimator. After undergoing diffraction at the apodization aperture and propagating for a distance up to the \Rx{}-clip, the beam presents flat amplitude and phase profiles.} 

\added{Possible apodization techniques were reviewed, and a scheme with non-refractive optics was chosen for being more easily manufacturable. A particle swarm optimization technique is used to explore the parameter space of possible aperture designs, and both starshade and sawtooth shapes are considered. The starshade shape (Figure~\ref{figure:APO-aperture}) is found to be the best performing shape. This aperture shape is known to yield an ``offset hyper-Gaussian'' transmission function $A(r)$, given in polar coordinates by~\cite{Cash2011},}
\begin{equation}
	A(r) = 1 \quad \text{for} \quad r < a,
\end{equation}
and
\begin{equation}
	A(r) = \exp \left(- \left[ \frac{r-a}{b} \right]^n \right) \quad \text{for} \quad r \geq a,
\end{equation}
\added{where $r$ is the radial distance from the center of the aperture, $a$ is the radius of the central region with total transmission, $b$ is the $1/e$ radius of the exponential function, and $n$ is the order of the hyper-Gaussian, determining how quickly the function falls with radius.

The shape of the aperture is optimized based on maximizing the flatness of the diffracted field pattern at the position of the \Rx{}-clip, which is calculated via a Fast Fourier Transform method. The boundaries of the parameter space of the optimization are tuned empirically to reduce the computation time, but are left large enough that a substantial range of possible designs are analyzed. The parameter space spanned the number of ``petals'' or teeth of the shape, the inner clear aperture radius, the order of the hyper-Gaussian, as well as the minimum width of the petal and the minimum width of the gap between adjacent petals. The theoretical transmission function never reaches zero, but becomes vanishingly small with radius as the width of the petals tends to zero. The machined part, obtained from a thin $\sim 100\,\mu$m metal foil using laser etching, is limited by a minimum gap size of 20\,$\mu$m (the resolution of the laser etching process, which arose from discussions with laser machining companies), which means that the real transmission function reaches zero at some distance from the center of the aperture.}

The fiber collimator tube is mounted on the TS via two rings glued with flexure feet to the baseplate. The mounting structure is thermally compensated by a dual aluminum-titanium assembly, and the center of the tube is kept at a constant height to reduce beam jitter. The structure has flexures in the longitudinal and the vertical direction to avoid mounting stress.

The lateral alignment accuracy of the aperture and the effect on the resulting wavefront was investigated. The performance of the system was obtained via numerical simulation assuming a lateral alignment accuracy of $\pm 250\,\mu$m \added{(which is easy to attain by a skilled person)} and a nominal peak-to-valley (PV) wavefront error of the collimated Gaussian beam of $\lambda/6$ over 10\,mm \added{(this figure comes from discussions with Sch\"after \& Kirchhoff)}. The resulting configuration yields an optical field with a flat wavefront at the position of the \Rx{}-clip. The resulting phase flatness is $\lambda$/100 PV nominally, or up to $\lambda$/25 PV including tolerance, which is extremely good considering the cost and simplicity of the device. The intensity flatness is not as good, sitting at 5.4\% over 3\,mm, or 11\% including tolerance. The power transmission through the 2.2\,mm diameter \Rx{}-clip is around 3\%.

\added{Having the ability to produce a tilting flat-top beam representative of the LA interferometer thanks to the aforementioned flat-top beam generator, as well as a tilting fundamental Gaussian beam representative of the TM interferometer as reported in~\cite{Trobs2018}, makes the TS an excellent optical ground support equipment (OGSE) candidate for aiding the alignment and characterization of the LOB.}

\subsection{Laser preparation}

A frequency doubled Nd:YAG laser based on a diode-pumped non-planar ring oscillator is locked to an iodine standard providing a 1064\,nm output with an estimated frequency noise of $300\,\text{Hz}/\sqrt{\text{Hz}}$. The source is split in a table-top laser preparation bench and passed through three acousto-optic modulators (AOMs) to generate the three frequency components needed for the \Rx{}, \Tx{} and \LO{} beams. Since the TTL coupling measurement is independent of frequency, we use heterodyne signals of few kHz, as opposed to MHz, for increased simplicity of the readout electronics. The \Tx{}-\Rx{}, \Tx{}-\LO{}, and \Rx{}-\LO{} beat notes are 9.765625, 14.6484375, and 24.4140625\,kHz respectively, chosen to avoid harmonic relations. These signals are captured by a pair of GAP1000Q InGaAs quadrant photodiodes with 1\,mm active area radius and 20\,$\mu$m slit width.

\subsection{Phase metrology system}

For precise readout of DC components, AC amplitudes, and phase of the heterodyne signals, a custom-designed Phase Measurement System (PMS) with 16 channels has been developed, with a phase noise performance of $\mu\mathrm{rad}/\sqrt{\mathrm{Hz}}$ in the 0.1\,mHz to 1\,Hz observation band. The DC component is proportional to the DC optical power on each QPD segment, and it is used to determine the beam positions using differential power sensing (DPS)~\cite{Killow2013}. Amplitude and phase readout are based on an in-phase/quadrature (I/Q) demodulator applied individually to each photoreceiver output channel and implemented onto a single field-programmable gate array (FPGA) for real-time operation at 80 MSPS~\cite{Heinzel2004}. 

The raw I/Q measurements, after filtering and downsampling, are delivered to an external computer for back-end processing, computing amplitude readout as the quadrature sum of both components and phase readout as the arctangent of quadrature over in-phase.  The phase of the coherent sum of all four QPD segments is used to measure longitudinal displacements with picometer precision, whereas the phases differences between pairs of QPD segments (i.e., the DWS signals) are used to compute horizontal and vertical angular signals between interfering beams with nanoradian precision. By combining the DWS and DPS signals, the control software computes a calibration matrix translating horizontal and vertical phase shifts into yaw and pitch angular tilts, respectively.

The internal clock of the PMS serves as the frequency reference to the RF signal generators that drive the AOMs, which reduces the differential clock noise between the heterodyne signals. Furthermore, the PMS is used to implement the \Rx{}-\LO{}, and the \Tx{}-\LO{} optical phase-locked loops using the single element photodiode (SEPD) in the TS. The OPLLs are based on a pair of piezo-driven mirrors placed in the optical paths of the \Rx{} and \Tx{} beams in the laser preparation bench before being fiber-coupled~\cite{Killow16AO} into the testbed. The \Rx{} beam can also be amplitude-modulated at 200\,Hz when required to aid with the alignment of this beam in the TS and OB independently of the other beams.

The pathlength noise due to the combined readout noise sources (shot noise, electronic noise, and digitization noise) is significantly below $1\,\mathrm{pm}/\sqrt{\mathrm{Hz}}$, given the laser power levels used in the testbed~\cite{Lieser2017PhD}. Since the investigation of TTL coupling does not require picometer stability, the testbed was operated in air and a temperature stability of $10^{-4}\,\mathrm{K}/\sqrt{\mathrm{Hz}}$ was reached using only passive thermal insulation. This translates into a pathlength noise floor of approximately $1\,\mathrm{nm}/\sqrt{\mathrm{Hz}}$, which gives rise to an effective error in the measurement of the TTL coupling that is estimated to be on the order of $0.5\,\mu$m/rad.

\section{TTL Coupling and Imaging Systems}
\label{section:imaging-systems}

\subsection{TTL coupling in the LA interferometer}

\begin{figure}[b]
\centering
	\includegraphics[scale=0.8]{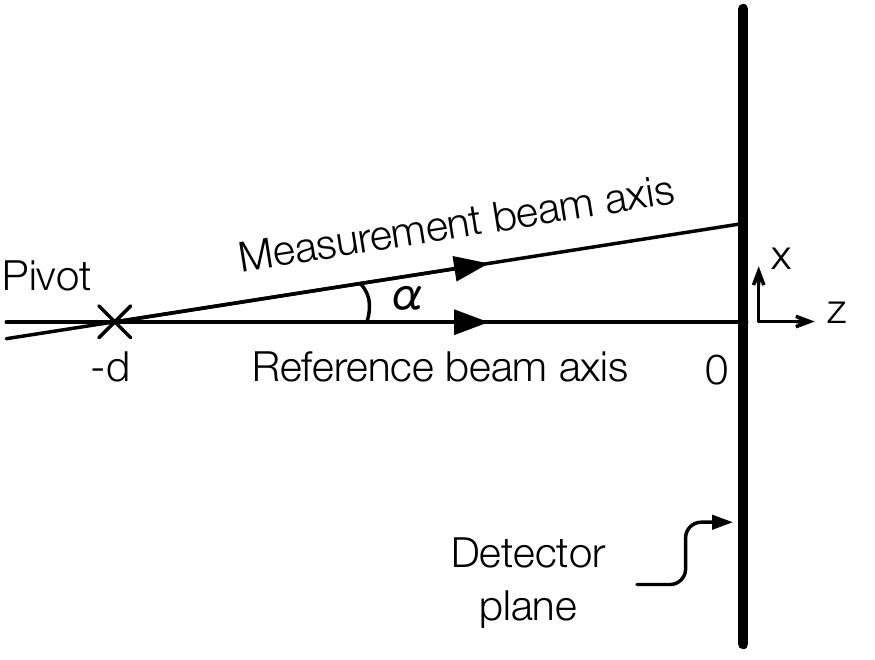}
	\caption{Lever arm tilt between two beams. Two beams (the ``reference'' and ``measurement'' beams) overlap. Their superposition is captured at a photoreceiver whose surface is the $z=0$ plane. The measurement beam rotates about the out-of-plane $y$-axis and the pivot point $(0,0,-d)^{\intercal}$. The tilt of the measurement beam couples into the longitudinal pathlength change sensed by the detector, which is a source of noise known as tilt-to-length coupling.}
	\label{figure:leverarm}
\end{figure}

\begin{figure}[b]
\centering
	\includegraphics[scale=0.8]{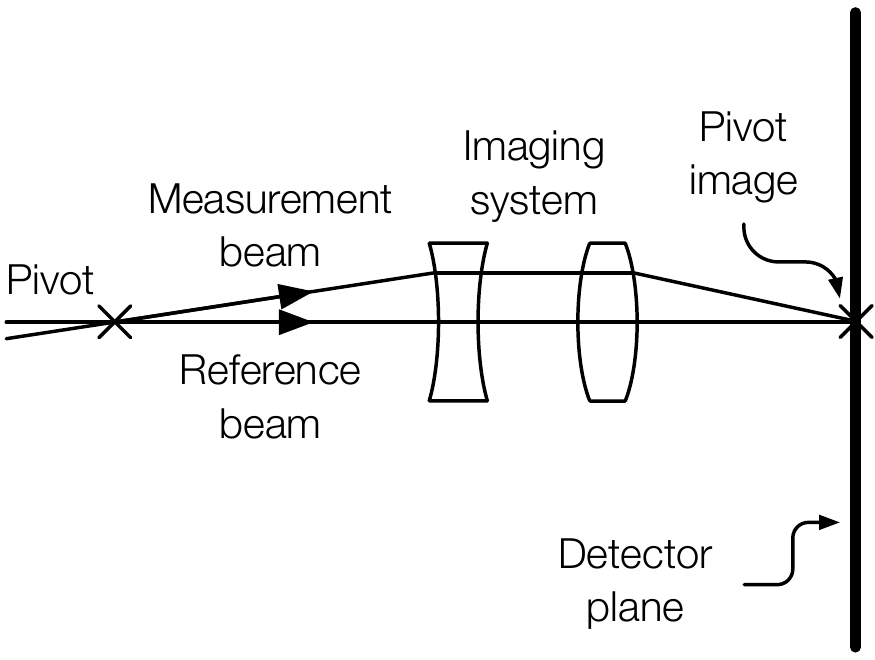}
	\caption{TTL coupling suppression via imaging. A set of lenses is configured to image the point of rotation of the measurement beam onto the center of the detector, so that the optical pathlength is unchanged by the tilt, reducing the geometric component of the coupling of the tilt to the longitudinal pathlength sensed by the detector. A contribution will remain due to the non-geometric component, which may be counteracted by tuning the position of the detector.}
	\label{figure:imaging-concept}
\end{figure}

Tilt-to-length coupling in two-beam interferometers is a complex effect. \emph{Geometric} TTL coupling originates when one of the beams, typically the beam carrying the main measurement signal, becomes tilted and deviates from the nominal interferometer topology, experiencing a longer propagation than the well aligned reference beam. The optical pathlength difference between the beams results in a phase shift that can be observed, e.g., as power fluctuations in a photodiode in a homodyne interferometer, or as a shift of the beat note's phase in a heterodyne scheme. 

For example, for a lever arm tilt as shown in Figure~\ref{figure:leverarm}, plane waves exhibit a geometric TTL coupling that can be derived from simple trigonometry to be $\frac{d}{2}\alpha^2$ to second  order in $\alpha$, where $d$ is the pivot length and $\alpha$ the tilt angle. However, in an interferometer with real beams, the particular features of the wavefront of the interfering beams at the position of the detector come into play. Hence, TTL coupling can be observed even if the optical pathlength is unchanged by the tilt (i.e., \emph{non-geometric} TTL coupling). For example, beam axes offsets, wavefront curvature mismatches or wavefront distortions, or, in general, effects disturbing the distribution of the phase of the overlapped optical fields in the detector surface, can be sources of non-geometric coupling.

The longitudinal pathlength signal $s_{\text{LPS}}$ can be computed as the complex phase of the integral of the overlapped optical fields over the active area of the detector~\cite{Wanner10PhD},
\begin{equation}
	s_{\text{LPS}} \equiv \frac{1}{k}\arg\left(\iint  E_m E_r ^{\ast} dA_{\text{pd}} \right).
	\label{equation:LPS}
\end{equation}
where $k=2\pi/\lambda$, $E_m$ is the measurement beam, and $E_r$ is the reference beam. Any dependence of $s_{\text{LPS}}$ with beam misalignment can be a source of TTL coupling. A closed expression for the LPS can be found for the case of two misaligned and mismatched Gaussian beams~\cite{Wanner2014, Schuster15AO} by considering a single-element photoreceiver with an active area radius much larger than the Gaussian radius of the beams. \added{The problem becomes more complex as we consider realistic detectors, such as a quadrant detector of finite size with gaps between segments. However, we have efficient and proven methods for computing interferometric signals (such as LPS, DPS, or DWS signals) for interferometers with two Gaussian beams~\cite{Ifocad, Wanner12OC}, and a plethora of experimental data.}

\begin{figure*}
\centering
	\includegraphics[scale=1]{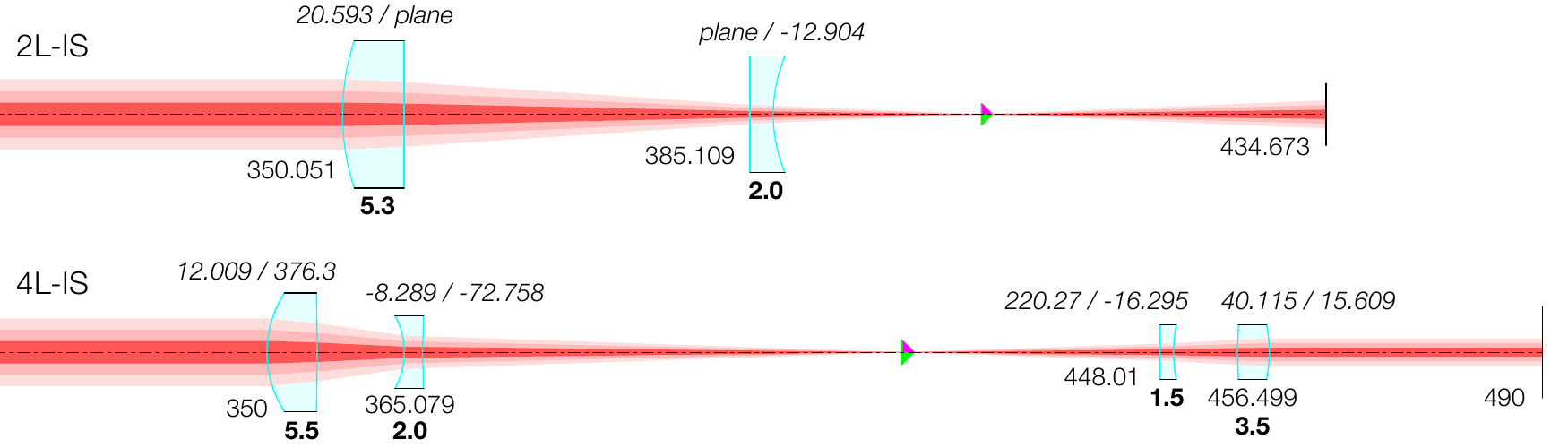}
	\caption{Imaging systems used in the experiment. The two-lens imaging system (2L-IS) is designed by means of a computer simulation which minimizes beam walk at the photodiode for a given amount of tilt, sweeping the parameter space of the lens choice and the lens positions. The four-lens imaging system (4L-IS) is designed using a traditional ray optics approach, by tuning the system's parameters to yield a ray transfer matrix of the form given in Equation~\ref{equation:imaging-system}. The values printed on the top of the lens in italic and separated by a dash `/' represent the radii of curvature of the primary / secondary surfaces of the lens (in millimeters$^{-1}$). The values printed on the bottom of the lens in bold represent the thickness of the lens in the optical axis (in millimeters). The values in regular font represent the lens and detector positions (in millimeters). The position of each lens is determined by the intercept of its primary surface with the optical axis.}
	\label{figure:imaging-systems}
\end{figure*}
 
\added{In the LA interferometer, however, the situation differs significantly, as the reference beam $E_r$ is Gaussian, but the measurement beam $E_m$ is a flat-top beam. This beam may be described, at the position of the \Rx{}-clip, as having a plane wave amplitude and phase in the region described by a disk of certain radius (determined by the receiving telescope parameters), and zero elsewhere. Moreover, the measurement beam suffers diffraction at the \Rx{}-clip aperture, and it is then imaged onto the detector plane. Unfortunately it is not possible to express Equation~\ref{equation:LPS} in terms of elementary functions, even in the infinite single-element detector approximation. There is ongoing work on developing both numerical and analytical methods for computing interferometric signals between a Gaussian beam and a flat-top beam using realistic detectors~\cite{Zhao2019}. In the scope of this work, however, we are only interested in the residual TTL coupling that arises from the interference between these beams in an interferometer with imaging systems.}

\subsection{Imaging systems to suppress TTL coupling}
\label{section:IS}

Since the measurement beam tilts around the \Rx{}-clip, imaging this plane onto the detector plane (Figure~\ref{figure:imaging-concept}) means that the geometric component of the cross-coupling of the tilt to the longitudinal signal is suppressed, leaving only the contribution of the non-geometric component, as well as a residual geometric coupling due to fundamental noise sources, such as thermal displacement noise. Hence, tilt-to-length coupling can be reduced greatly by placing specially tuned imaging systems in front of the photoreceivers~\cite{Schuster16OE}.

\added{In this paper we describe the performance of two different imaging systems for TTL coupling reduction in the LISA long-arm interferometer, a two-lens imaging system (2L-IS), and a four-lens imaging system (4L-IS), as shown in Figure~\ref{figure:imaging-systems}. Both systems have been already used in~\cite{Trobs2018} in the context of the test mass interferometer. In this section we present the different methodologies that were followed to design each system.}

The 4L-IS was designed using a classic pupil-plane imaging system approach. The system is configured so that rays $(x, y, \alpha, \beta)^{\intercal}$ originating from the \Rx{}-clip are imaged onto the detector plane with a certain magnification, $(x',y') = m (x, y)$, where $(x, y)$ is a point within the \Rx{}-clip aperture, $m$ is the lateral magnification, and $(x', y')$ is a point at the detector surface. This condition can be expressed using the \emph{ray transfer matrix} formalism as
\begin{equation}
\*r' = \begin{pmatrix} m & 0 \\ c & 1/m  \end{pmatrix} \*r,
\label{equation:imaging-system}
\end{equation}
where $\*r = (x, \alpha)^{\intercal}$, and we have omitted the $y$ and $\beta$ degrees of freedom due to the existing cylindrical symmetry. Note that in such system $x' = m x$ is $\alpha$-invariant (i.e., rays originating at the center of the \Rx{}-clip are projected onto the center of the photodiode). This is the fundamental condition for imaging. Moreover, the system was designed to produce a collimated output given a collimated input (i.e., $c = 0$), which makes the system more robust against beam parameter variations, and advanced techniques were used to minimize aberrations, ghosts, and the sensitivity of the system to manufacturing tolerances.

The 2L-IS was designed via numerical simulation using the interferometer modelling software IfoCAD~\cite{Ifocad, Wanner12OC}, and an algorithm which looks for solutions that yield very low beam walk at the detector plane for a given amount of beam tilt and a desired magnification. The simulation consists of an optical setup such as the one described in Figure~\ref{figure:imaging-concept}, introducing two lenses between the pivot point and the photodiode. The algorithm sweeps through the parameter space spanned by the position and orientation of the lenses, as well as the choice of the lenses themselves. Both systems were designed to use commercially-available lenses from a popular lens manufacturer. \added{The program computes the image of a test beam tilted by 100\,$\mu$rad with respect to the nominal optical axis. It is verified that the solutions do not change appreciably from run to run or by changing the tilt angle. The program can be scaled for designing imaging systems with any number of lenses with minor effort simply by increasing the size of the parameter space used in the optimization. The reason for using only two lenses in the design of 2L-IS is two-fold: first, to show that it is possible to achieve near-perfect performance in terms of raw TTL coupling suppression by employing only two lenses; second, to design a system that is relatively more compact than 4L-IS, which helps deal with surface space constraints on the optical bench. }

\begin{figure*}
\centering
	\includegraphics[scale=1]{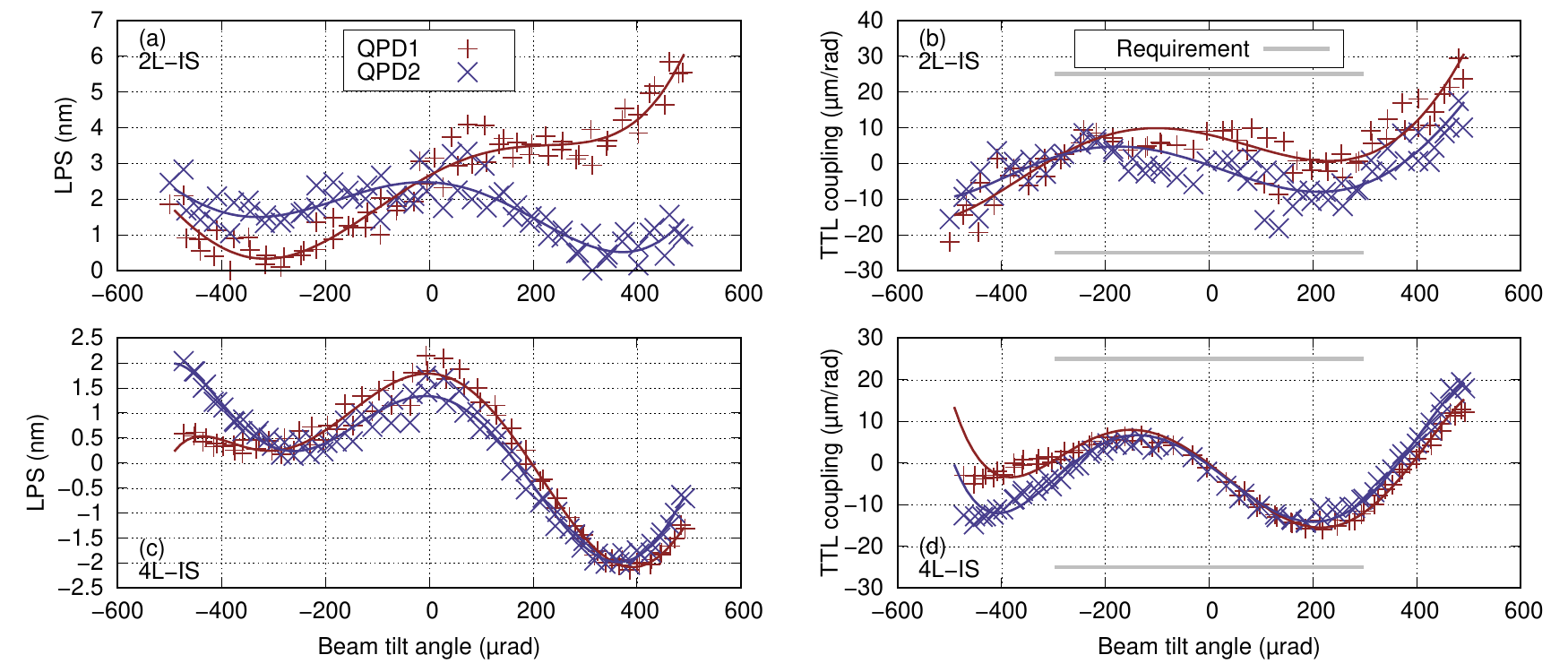}
	\caption{\added{Optical suppression of tilt-to-length coupling in the long-arm interferometer with imaging systems, as sensed by the quadrant photodiodes at the output ports of the recombination beamsplitter in the measurement interferometer (QPD1 and QPD2, see Figure~\ref{figure:setup}). The telescope simulator is used to induce a lever arm tilt of the measurement flat-top beam with respect to the stable reference Gaussian beam, so as to simulate angular jitter of the spacecraft in the LISA long-arm interferometer. The resulting longitudinal pathlength signal (LPS) and its first derivative (TTL coupling) are plotted vs the beam tilt angle, showing that both the two-lens (a, b) and the four-lens (c, d) imaging systems (2L-IS and 4L-IS) perform to the required level on an optimally aligned interferometer. The trend curves are obtained by fitting sixth order polynomials to the data.}}
	\label{figure:nominal-performance}
\end{figure*}

\added{Note that the solutions provided by the automated approach do not necessarily fulfill the imaging system condition as described by Equation~\ref{equation:imaging-system}. The resulting system's ray transfer matrix $M_{ij}$ does not verify $M_{12}=0$ exactly, as the algorithm may find an optimal solution that is close to but does not necessarily match this condition (i.e., some tolerance is allowed).} But the main difference between the two imaging systems considered here is that the two-lens system does not provide a collimated output beam for a collimated input beam. The 2L-IS is known to be more sensitive to changes in the beam parameters~\cite{Trobs2018, Schuster17PhD}, although it is more robust to lateral misalignments of the components of the system, as it is shown in Section~\ref{section:results-tolerance}.

\added{The use of free-form lenses was investigated in~\cite{Schuster17PhD}, with a modified version of the computer program used to design the two-lens system described here. In terms of raw TTL coupling suppression, assuming a perfectly aligned imaging system, both 2L-IS and 4L-IS offer near-perfect performance, and free form lenses do not provide a significant advantage. Other performance figures, such as the sensitivity to alignment errors of the system, could conceivably be improved, as reported in~\cite{Schuster17PhD}.} 

The imaging systems are mounted on the OB using an optomechanical subassembly~\cite{PerreurLloyd15JPCS} that allows for precision multi-axis adjustment of the lens positions. The optical mounts of the lenses and the detectors are based on thermally stable monolithic aluminium flexures with ultra-fine precision screws allowing for $1\,\mu$m positional precision and $<1\,$arc-minute angular precision.

\section{Results}
\label{section:results}

\begin{figure*}
\centering
	\includegraphics[scale=1]{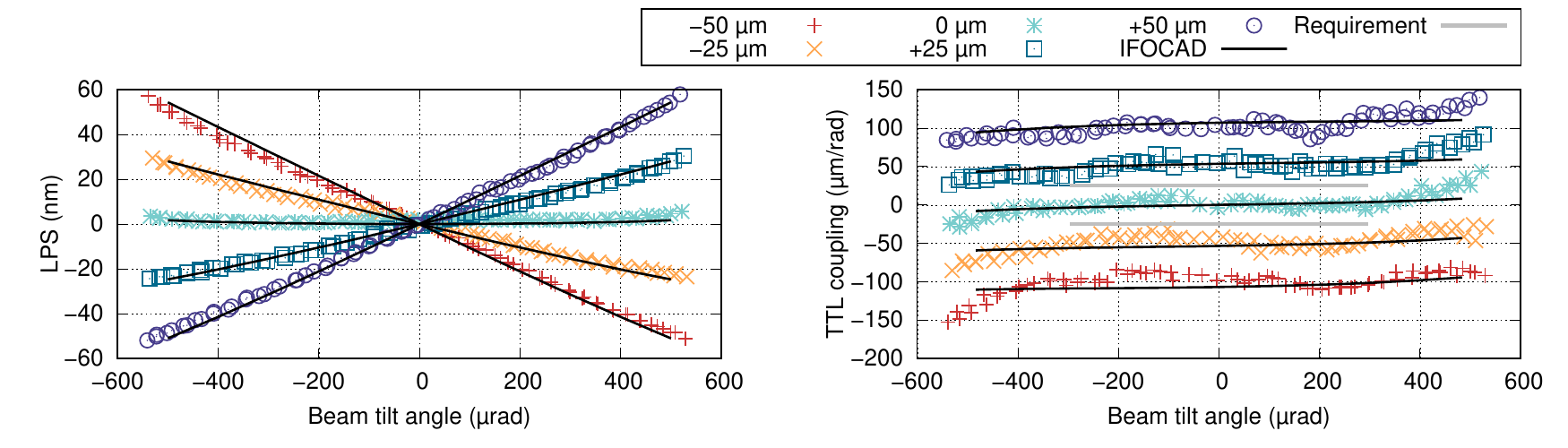}
	\caption{Lateral shifts of the detector. Shifting the detector vertically yields a linear dependency of the longitudinal pathlength signal with the beam tilt angle and, thus, an approximately constant value in its first derivative. For example, a shift of $+25\,\mu$m of the detector translates into a TTL coupling of approximately $+50\,\mu$m/rad. The measurements are compared against numerical simulations of the experiment using IfoCAD, showing good agreement considering the precision of the alignment screw that is used to shift the QPD.}
	\label{figure:2L-IS-lateral-QPD-offset}
\end{figure*}

In this section we report on the best achieved performance of the two imaging systems introduced in the previous section, as well as on their robustness to alignment deviations. For each imaging system we measure the resulting TTL coupling between the flat-top measurement beam and the Gaussian reference beam in the two quadrant photodiodes \added{(QPD1 and QPD2)} of the main measurement interferometer of the OB by introducing intentional tilts via the TS. The TS actuates on the measurement beam in a way that produces an out-of-plane tilt with respect to the \Rx{}-clip in the optical bench (i.e., the \Rx{} beam's propagation vector acquires a vertical component), and the relative phase between the measurement and reference beams is kept at zero at this point via the OPLLs implemented using the reference interferometer in the TS. 

The longitudinal pathlength signal $s_{\text{LPS}}$ is obtained as the average phase over the four QPD segments,
\begin{equation}
	s_{\text{LPS}}^{\text{AP}} = \frac{\psi_{\text{A}}+\psi_{\text{B}}+\psi_{\text{C}}+\psi_{\text{D}}}{4k},
\end{equation}
where $\psi_{\text{A}}$ refers to the phase measured by the QPD quadrant A, and so forth. This is measured as the tilt actuators on the TS sweep the beam tilt angle. The TTL coupling factors are then computed as the first derivative of this signal with respect to the tilt angle.

Prior to tilting the beam, the actuators find the position of the \Rx{} beam's nominal axis. This is attained in three steps: first, a rough initial alignment is performed by commanding the actuators to move to an absolute position where contrast can be observed in the reference QPD (the amplitude-modulated \Rx{} signal is useful for this); then, the actuators find the position which maximizes the amplitude of the heterodyne beat note between the \Rx{} and \LO{} beams in the reference QPD; lastly, the actuators acquire a position which minimizes the DWS signal, also in the reference QPD. This procedure is performed for both axes. After the \Rx{} beam has been successfully aligned, the control software commands the actuators to tilt the beam from zero to the minimal and then to the maximal tilt angles. For each step within the tilt range, the angle is measured using the calibrated DWS signals from the reference QPD in the TS, and the LPS is measured by averaging 900 samples. The TLL coupling (i.e., the first derivative of the LPS) is then computed for each data point via a piecewise linear regression of the five neighbouring points (i.e., $i$, $i \pm 1$, $i \pm 2$). 

\subsection{Nominal performance}

\added{We first report on the nominal performance of both systems (Figure~\ref{figure:nominal-performance}), i.e., on the measure of the achieved performance after the alignment of the system has been optimized and the TTL coupling reduced to our best effort. After the imaging systems have been installed and aligned in the testbed, the goal is to find the optimal detector position that yields the least amount of TTL coupling. We distinguish between transverse and longitudinal offsets from the optimal detector position by looking at the resulting TTL coupling dependence with tilt angle; transverse offsets yield a linear dependence, while longitudinal offsets yield a quadratic dependence.}

\added{This optimization is performed in three steps. The first step is tuning the longitudinal position of the QPDs to find the rough position of the exit pupil of the imaging systems, at which point the detectors sense no beam walk even if the measurement beam is tilting, and hence, geometric TTL coupling is suppressed. At this point, however, TTL coupling suppression is not necessarily optimal, and we measure some coupling due to non-geometric effects. To reach the performance shown in Figure~\ref{figure:nominal-performance}, two additional steps of optimization are needed: first, the QPD is laterally offset to the center of the tilting measurement beam by finding a zero crossing in the DPS signal, so as to find very precisely the point in the plane where the center of rotation of the flat-top beam is being imaged; then, an additional lateral offset from this point is performed in order to further counteract the residual coupling. The optimum performance of the system is therefore obtained by balancing a non-geometric TTL coupling component with an intentionally imparted geometric component.}

\added{After performing these steps, both systems are able to fully meet the required specification of TTL coupling factors at or below $\pm 25\,\mu$m/rad for beam tilt angles within $\pm 300\,\mu$rad, a similar performance to what was found when the testbed was operated in the test mass interferometer configuration~\cite{Trobs2018}}. The two-lens system exhibits slightly larger deviations from the trend, an effect that is not attributed to any differences in the system design but rather correlates with the changing environmental conditions of the testbed at the time of the measurement. The remaining TTL coupling is at or below the noise floor of the testbed, and thus a further reduction would not yield new information about the performance of the imaging systems. 

\begin{figure*}
\centering
	\includegraphics[scale=1]{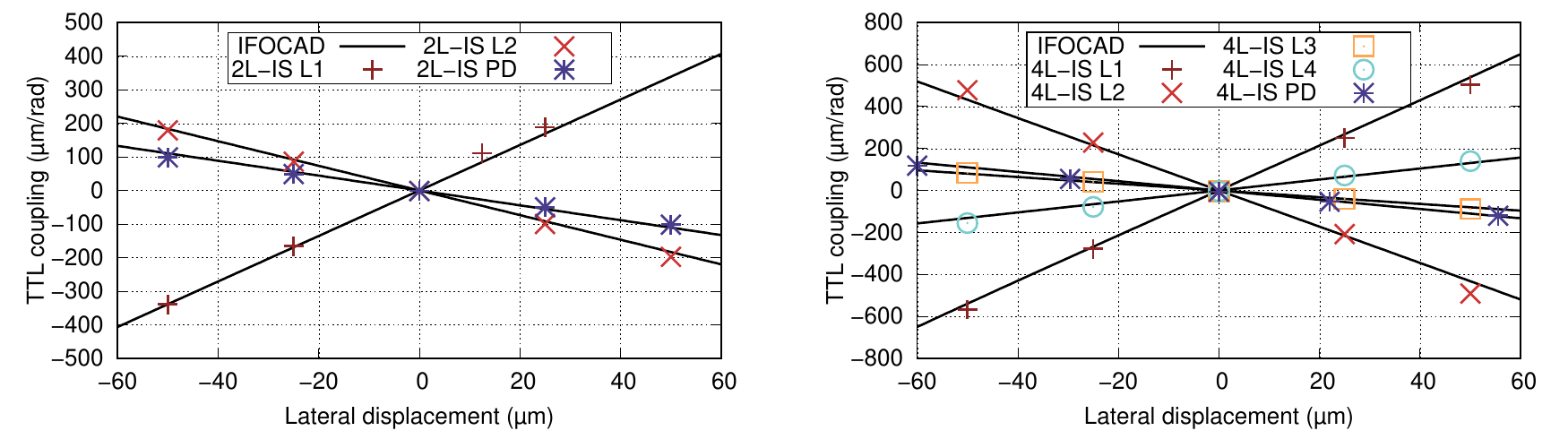}
	\caption{Sensitivity of the interferometer to lateral shifts of the different components of the two-lens (2L-IS) and the four-lens (4L-IS) imaging systems (e.g., L1 = first lens, L2 = second lens, PD = photodetector etc). These measurements, which are derived from the slope of a linear regression of the LPS vs beam tilt angle data, agree well with the IfoCAD model of the experiment. The two-lens system offers greater robustness against alignment errors, yielding a lower TTL coupling factor for a given shift of the lenses. }
	\label{figure:tolerance-analysis}
\end{figure*}

\subsection{Tolerance analysis}
\label{section:results-tolerance}

Having investigated the nominal performance of each imaging system, we carry out an investigation of the robustness of each system against alignment errors of the components, as well as of the position of the systems relative to the testbed, in order to determine the setup tolerances. \added{This helps us establish requirements for the manufacture and deployment of imaging systems, as well as discriminate between different imaging system implementations, or diagnose a non-optimally aligned system.}

The analysis is performed by introducing intentional misalignments in the testbed and measuring the resulting TTL coupling. The measurements also offer a good opportunity to verify our simulation tools by comparing the obtained results with an IfoCAD model of the experiment. The model includes the components specified in Figure~\ref{figure:imaging-systems}. The reference beam is treated as a general astigmatic Gaussian beam with a 1\,mm radius waist located at the pivot point (i.e., located at $(0,0,-d)^{\intercal}$ as indicated in Figures~\ref{figure:leverarm} and~\ref{figure:imaging-concept}). The reference beam parameters were measured in~\cite{Trobs2018}.

IfoCAD does not yet include solidly proven methods for simulating real flat-top beams. However, for a situation in which the \Rx{}-clip is imaged onto the detector plane, the beam may be modelled as a Gaussian beam set to yield a nearly flat intensity and phase profile within the detector's active area. The detector is a quadrant photodiode with 1\,mm active area diameter and $20\,\mu$m slit width.

Several degrees of freedom were investigated for both the two-lens and the four-lens systems. For example, Figure~\ref{figure:2L-IS-lateral-QPD-offset} shows the sensitivity of the two-lens imaging system to lateral shifts of the QPD in the vertical direction. The measurements show that a lateral misalignment of $\pm 25\,\mu$m results in an additional TTL coupling of $\approx\pm 50\,\mu$m$/$rad, and this is verified by simulation. This is in good agreement with the expected extra TTL coupling resulting for a lateral shift of the detector in the direction normal to the axis of rotation of the tilting beam, which is directly proportional to the magnitude of the shift. In the presence of an imaging system, this extra coupling is scaled by $1/m$, where $m$ is the magnification of the imaging system. The small deviation found here (\added{$m=0.4$} for the 2L-IS, so the expected geometric coupling factor is $\approx 2.5$) could be due to the precision of the alignment screw that is used to adjust the lateral shift of the QPD, or an additional non-geometric effect not accounted for. The alignment screw has a pitch of $200\,\mu$m per turn and thus, e.g., a misalignment of $25\,\mu$m  has an associated uncertainty of a few micrometers.

Similar investigations are carried out for all critical parameters of the system, namely the lateral alignment of each lens and the QPDs. An overview of the different tests is presented in Figure~\ref{figure:tolerance-analysis} for both imaging systems. Since the LPS is clearly linear in $\alpha$, each point in Figure~\ref{figure:tolerance-analysis} is obtained as the slope of a linear regression of the LPS for a given parametric misalignment. The results are in good agreement with the IfoCAD model of the experiment, while the small deviations are likely due to the precision to which the alignment errors are estimated.

We find that the two-lens system is in general more robust to alignment errors than the four-lens system. For the four-lens system the more critical parameters are the first and second lenses, showing a maximal TTL coupling in the range of $\pm 700$\,$\mu$m/rad for lateral shifts of 60\,$\mu$m. In other words, the 4L-IS shows an effective TTL coupling sensitivity of $11.5\,$rad$^{-1}$ to L1 and L2 alignment errors. This is in contrast with the third and four lenses of the system, both having associated sensitivities below $3.3\,$rad$^{-1}$. In the two-lens system, the sensitivities are $7.5\,$rad$^{-1}$, $4.2\,$rad$^{-1}$, and $2.5\,$rad$^{-1}$ for the first lens, second lens, and photodiode respectively.

\subsection{Compensation by photodiode alignment}

\begin{figure*}
\centering
	\includegraphics[scale=1]{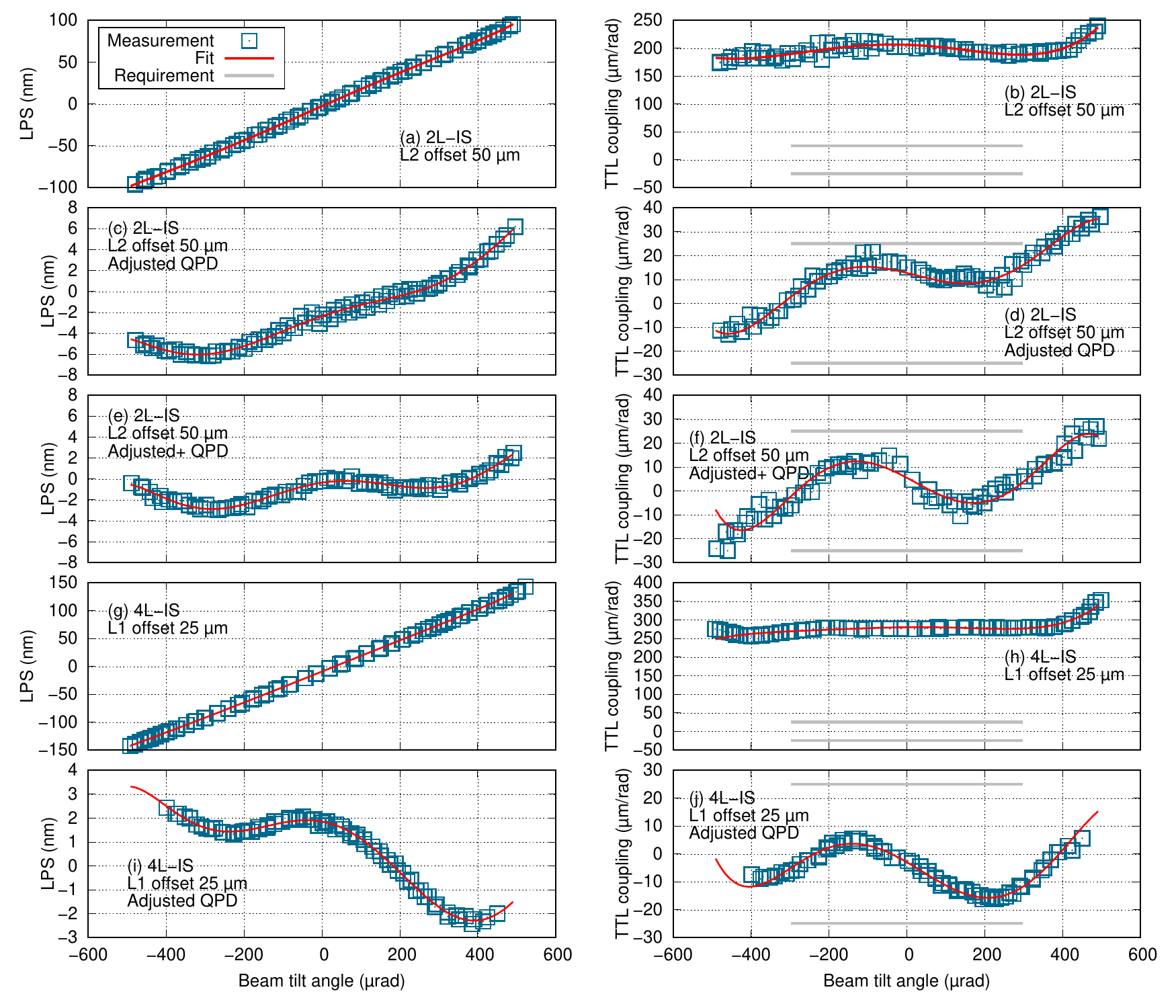}
	\caption{Compensation of misaligned imaging systems by photodiode alignment. Plots (a-b) show the results of a misaligned second lens in the two-lens imaging system, and plots (c-d) and (e-f) show the compensated system that is achieved by vertical adjustment of the QPD position (the former by centering the flat-top beam in the QPD, and the latter by slightly shifting the QPD from the center of the flat-top so as to minimize the coupling). Plots (g-h) and (i-j) show the same situation for a misaligned first lens in the four-lens imaging system and the achieved compensation respectively. The trends are shown by fitting sixth order polynomials to the data.}
	\label{figure:photodiode-compensation}
\end{figure*}

We have shown how the interferometer responds to alignment errors in the imaging systems by intentionally misaligning parts of the system. In this section we investigate how an existing misalignment can be counteracted by laterally shifting a different component. In some situations the system may not perform or be configured as expected (e.g., due to manufacturing imperfections or shock). However, in some situations, the system may be reconfigured to achieve the required performance by introducing a lateral offset of another component. For example, in an interferometer with a misaligned imaging system yielding a large TTL coupling, it is often possible to compensate the system by introducing an additional lateral shift of the detector.

For example, for an alignment error of $50\,\mu$m of the second lens in the two-lens imaging system, which yields the linear dependency of the LPS with the beam tilt angle shown in Figure~\ref{figure:photodiode-compensation}\,a-b with a TTL coupling of $\sim 200\,\mu$m/rad, a realignment of the QPD to the center of the tilting measurement beam (e.g., by finding a zero crossing in the DPS signal) significantly reduces TTL coupling and returns the system to expected performance (Figure~\ref{figure:photodiode-compensation}\,c-d). Moreover, a further reduction is possible by slightly shifting the QPD from the center of the measurement beam so as to minimize the residual coupling (Figure~\ref{figure:photodiode-compensation}\,e-f). The same approach applies to alignment errors in the four-lens imaging system. Figure~\ref{figure:photodiode-compensation}\,g-h and i-j show the effect of a vertical shift of $25\,\mu$m of the first lens and the respective improvement by realignment of the photodiode to the center of the measurement beam, respectively.

This procedure was actually performed in order to reach the nominal performance shown in Figure~\ref{figure:nominal-performance} for both the 2L-IS and 4L-IS systems. The imaging systems were pre-aligned during manufacturing using a different setup. In the testbed, however, it is not possible to control the beam height with precision, and thus the imaging systems were by design laterally offset from the nominal interferometer topology. Implementing the compensation of the imaging systems by photodiode alignment was therefore a critical step in order to reach performance.

\section{Summary and conclusion}
\label{section:conclusion}

The coupling of angular noise into the longitudinal pathlength readout is an aspect of utmost importance in space interferometers such as LISA Pathfinder, GRACE-FO, and LISA. This tilt-to-length coupling noise is considered in this paper in the context of the LISA long-arm interferometer. We have demonstrated the use of imaging systems to reduce TTL coupling in a setup representative of the LISA long-arm interferometer, a major step towards validating this noise-reduction strategy for LISA. \added{The use of such systems is the current baseline for passive optical reduction of TTL coupling in the LISA Optical Bench.}

For this purpose an ultra-stable interferometer testbed was built, and its operation is described in Section~\ref{section:testbed}. The testbed was used previously to test the performance of the imaging systems in a configuration representative of the LISA test mass interferometer~\cite{Trobs2018}. The testbed is a simplified version of the LISA Optical Bench, with all the components required for a TTL coupling investigation, and features a telescope simulator. \added{For experiments in the long-arm interferometer configuration, the testbed is fitted with a flat-top beam generator that provides a laser beam with a flat intensity and phase profiles simulating the beam received by a LISA spacecraft. The shape of this measurement beam makes this interferometer unique.} The telescope simulator developed for this experiment has proven to be a powerful optical ground support equipment candidate for the LISA Mission. A similar device may be used to aid alignment of the LISA Optical Bench and to calibrate the TTL coupling during construction.

It was shown experimentally for the two-lens and the four-lens imaging systems described in Section~\ref{section:IS} that the TTL coupling in the testbed could be reduced below the level of $\pm$25\,$\upmu$m/rad for beam tilt angles within $\pm$300\,$\upmu$rad (Section~\ref{section:results}). Furthermore, we performed a tolerance analysis of both imaging systems and investigated the additional TTL coupling due to lateral alignment errors of each system. The results obtained demonstrate that TTL coupling can be counteracted by introducing intentional lateral shifts of other components in the system. For example it was shown that the lateral position of the detectors could be shifted as a compensation mechanism. These findings pave the way towards the development of advanced strategies for optical TTL coupling noise reduction in LISA.


\vspace*{1cm}



\begin{acknowledgments}

We acknowledge funding by the European Space Agency within the project ``Optical Bench Development for LISA'' (22331/09/NL/HB), support from UK Space Agency, University of Glasgow, Scottish Universities Physics Alliance (SUPA), and support by Deutsches Zentrum f\"ur Luft und Raumfahrt (DLR) with funding from  the Bundesministerium f\"ur Wirtschaft und Technologie (DLR project reference 50 OQ 0601). We thank the German Research Foundation for funding the cluster of Excellence QUEST -- Centre for Quantum Engineering and Space-Time Research. We acknowledge financial support of Deutsche Forschungsgemeinschaft (DFG) in the frame of SFB1128 geoQ, project A05 for the optical simulations.

\end{acknowledgments}



%


\end{document}